\documentclass[aps,twocolumn,prd,preprintnumbers,amsmath,amssymb,superscriptaddress,nofootinbib]{revtex4-1}

\usepackage[export]{adjustbox}
 \usepackage{color}
 \usepackage{epsfig}
 \usepackage{ulem}

\definecolor{White}{rgb}{1,1,1}
\definecolor{Red}{rgb}{1,0.1,0}
\definecolor{LightYellow}{rgb}{1,1,.875}
\definecolor{SteelBlue}{rgb}{.273,.508,.703}
\definecolor{navy}{rgb}{0,0,.5}
\definecolor{LightCyan}{rgb}{.875,1,1}
\definecolor{DarkRed}{rgb}{.543,0,0}
\definecolor{HotPink}{rgb}{1,.41,.70}
\definecolor{ForestGreen}{rgb}{.13,.54,.13}
\definecolor{OliveDrab}{rgb}{.42,.55,.14}
\definecolor{MediumBlue}{rgb}{0,0,.80}
\definecolor{RoyalBlue}{rgb}{.25,.41,.88}
\definecolor{DeepSkyBlue}{rgb}{0,.746,1}
\definecolor{Brown}{rgb}{0.545,0.271,0.074}
\definecolor{Purple}{rgb}{0.637,0.285,0.641}

\newcommand{\dis}[1]{\begin{equation}\begin{split}#1\end{split}\end{equation}}
\def\bea{\begin{eqnarray}}
\def\eea{\end{eqnarray}}

\def\L{{\cal L}}
\def\S{{\cal S}}
\def\R{{\cal R}}

\begin{document}

\preprint{PNUTP-23/A01} 
\preprint{CTPU-PTC-23-05} 
\preprint{APCTP Pre2023 - 002} 
 
\title{Holographic light dilaton at the conformal edge}

\author{Jes\'us Cruz Rojas}
\email{jesuscruz.rojas@apctp.org}
\affiliation{Asia Pacific Center for Theoretical Physics,   Pohang 37673,   Korea }
\author{Deog Ki Hong}
\email{dkhong@pusan.ac.kr}
\affiliation{Department of Physics,  Pusan National University, Busan 46241,   Korea }
\author{Sang Hui Im}
\email{imsanghui@ibs.re.kr}
\affiliation{Particle Theory  and Cosmology Group, Center for Theoretical Physics of the Universe,   Institute for Basic Science, Daejeon 34126,   Korea }
\author{Matti J\"arvinen}
\email{matti.jarvinen@apctp.org}
\affiliation{Asia Pacific Center for Theoretical Physics,   Pohang 37673,   Korea }
\affiliation{Department of Physics, Pohang University of Science and Technology,   Pohang 37673,   Korea }

\begin{abstract} 

We study a simple holographic model for gauge theories near the conformal edge to show that the dilaton can be parametrically lighter than any other composite states. The masses of all composite states, except the Nambu-Goldstone bosons like dilaton, are bounded by the infrared scale or the dynamical mass. The parametric dependence of the dilaton mass is controlled by the closeness of the anomalous dimension of the quark bilinear, that breaks spontaneously the scale symmetry, to the conformality.  We also show in the holographic dual that under certain assumptions, the dilaton saturates at low energy the anomalous Ward identity for the dilatation currents.

\end{abstract}  

\maketitle
\section{Introduction}

Why the Higgs mass is so light compared to the fundamental scale of elementary particles, like the Planck or GUT scales, has been one of the driving questions for physics beyond the standard model, as the Higgs mass in the standard model is seemingly extremely fine-tuned. Much of interest on naturally light Higgs has been focused recently on dynamically generating Higgs mass from quantum criticality or conformal edge, at which the BSM physics enjoys the scale invariance at high energies (UV)~\cite{Hong:2017smd,Eroncel:2018dkg}. This approach of explaining Higgs mass from conformal edge turns out to be  quite attractive in view of current experimental status of not finding any hint of new particles around the weak scale, $\Lambda_{\rm ew}\sim 1~{\rm TeV}$, that has been explored vigorously at current energy frontiers. 

If an infrared scale is dynamically generated from the quantum criticality, however, the modulus of the scale-invariant theory, so-called dilaton, naturally develops a mass as well. If the dilaton is sufficiently light, it can be a good candidate for dark matter~\cite{Choi:2011fy,Choi:2012kx}.

Another context where nearly conformal, quantum critical dynamics has received a lot of attention recently has been  complex conformal field theories (CFTs)~\cite{Gorbenko:2018ncu}. The basic idea is that two fixed points merge and move to complex values of the couplings, and a nearly conformal flow arises when the imaginary parts of the couplings are small. Such a picture is expected to appear also in QCD at low energies (IR) as one varies the number of flavors, near the so-called conformal window~\cite{Kaplan:2009kr}. Complex CFTs have been studied recently by using the gauge/gravity duality, or holography for short, in~\cite{Faedo:2019nxw,Faedo:2021ksi} (see also~\cite{Alanen:2010tg}). 

In this article we show from the holographic analysis that the dilaton from the quantum criticality can be quite light for certain low-energy parameters governed by the UV completion, compared to the dynamical mass, as long as the back reaction of dynamical mass does not perturb the criticality too much, keeping the theory near the conformal edge until reaching close to the infrared.

\section{A holographic description of the near conformal edge}

The AdS/CFT correspondence \cite{Maldacena:1997re} suggests that
a 4-dimensional (4D) Yang-Mills gauge theory with approximate conformal invariance
may have a holographic dual description in terms of a corresponding 5-dimensional (5D) gravity theory in approximate Anti-de-Sitter (AdS) geometric background. For 4D gauge theory with $N_c$ number of colors  and $N_F$ number of flavors, a simple 5D action containing proper modeling of both the gauge theory and matter sectors is given by
\dis{
\S_5=   \frac{1 }{2\kappa_5^2}  \int d^4 x \int dz \sqrt{g}\left( \R + \Lambda_5 + \L_5 \right),
}
where $1/(2\kappa_5^2) \equiv M_5^3$ with the 5D Planck mass $M_5$, 
$\Lambda_5 (>0)$ is a 5D cosmological constant, and
\dis{
\L_5\! = \!\textrm{Tr} \left[| D_M X |^2 - M_X^2 |X|^2\right] -\sum_{i=L,R}\textrm{Tr}\left[ \left(F_i^{MN}\right)^2\right] 
}
with $D_M$ being the covariant derivatives and the indices $M, N$ running over $(\mu, 5)$, while the Greek index $\mu$ runs over $(0,\dots, 3)$. 
Here $X$ is a 5D complex scalar transforming as bi-fundamental $(N_F, \bar{N}_F)$ under the bulk gauge symmetry ${\rm SU}(N_F)_L \times {\rm SU}(N_F)_R \times {\rm U}(1)_B$, and $F_L^{MN}$ ($F_R^{MN}$)  are the field strength tensors of the corresponding bulk gauge fields.
The 5D scalar field $X$ provides a dual description of the 4D quark bilinear operators $\bar{q}_L^i q_R^j\, (i,j = 1, \dots, N_F)$, and the gauge fields are dual to the quark currents $\bar{q}_L^i \gamma^\mu q_L^j$ and $\bar{q}_R^i \gamma^\mu q_R^j$, respectively. 

Neglecting for a moment the effect of the matter sector to the geometry, the approximate background 5D metric is given by the AdS$_5$ geometry, which may be written in conformal coordinates as
\dis{
ds^2 = a(z)^2 \left( \eta_{\mu \nu} dx^\mu dx^\nu - dz^2 \right) \label{AdSm}
}
where $\eta_{\mu \nu} = (1, -1, -1, -1)$, and the warp factor $a(z)$ is given by
\dis{
a(z) = \frac{L}{z} \label{warp}
}
with $L^2 = 12/\Lambda_5$. A tiny 4D conformal symmetry breaking is described in the 5D dual theory by a little departure from the AdS$_5$ geometry induced by a non-zero vacuum expectation value (VEV) of $X$. 
The bulk scalar field $X$ can develop a small non-zero VEV when its mass squared $M_X^2$ is slightly below the Breitenlohner-Freedman (BF) bound $M_X^2 = -4/L^2$ so that
\dis{
M_X^2 = -\frac{4+\epsilon}{L^2}\,
}
with $0< \epsilon \ll 1$. According to the AdS/CFT correspondence, 
the 5D mass squared $M_X^2$ is related to the scaling dimension of the 4D quark bilinear operator $\bar{q}_L q_R$:
\dis{
\textrm{Dim}[\bar{q}_L q_R] = 2 + \sqrt{4 + M_X^2 L^2}.
}
Thus the quark bilinear operator develops a complex dimension for $\epsilon > 0$ signaling the loss of conformality~\cite{Kaplan:2009kr,Gorbenko:2018ncu}.

The standard picture is that the coupling of the gauge theory runs into a fixed point at low energies (IR) when $N_F$ is above a critical value, where the fixed point value of the quark bilinear operator hits two, and the slight violation of the BF bound takes place when $N_F$ is right below the critical value~\cite{Appelquist:1996dq,Kaplan:2009kr}. At high energies, 
due to the asymptotic freedom, the dimension of the operator flows to the free value, three. Here we will ignore the flow and focus on the physics near the fixed point, taking $M_X$ to be constant. This approach is enough to capture walking, large separation between the IR and UV energy scales~\cite{Hong:2006si}, and allows us to obtain most of the results analytically.

For the background geometry (\ref{AdSm}), the equation of motion of $X$ is given by
\dis{
 X'' + 3\frac{a’}{a}  X’ - a^2 M_X^2 X - \square_4 X=0,
}
where the prime denotes derivatives with respect to the coordinate $z$, and $\square_4 \equiv \eta^{\mu \nu} \partial_\mu \partial_\nu$. For the zero mode $\square_4 X=0$, the background solution for $X$ with the warp factor (\ref{warp}) 
may be written as
\dis{
\bar{X}(z)= \sigma z^2 \sin \left(\sqrt{\epsilon} \ln \left(\frac{z}{z_{\rm UV}} \right) + \alpha \right), \label{X0}
} 
where the parameter $\sigma$, corresponding to the quark condensate evaluated at UV, has mass dimension 2 and $0< \alpha < \pi$. The parameter $\alpha$ depends on the physics above the UV cutoff, $0<z<z_{\rm UV}$. Let us assume that the extra dimension is cut off in the infrared (IR), at $z=z_{\rm IR}$, which can be interpreted as generation of a dynamical scale $m_{\rm dyn} \sim z_{\rm IR}^{-1}$, explicitly breaking the scale symmetry.  The background solution should not cross the nodes of the sine function in $ z_{\rm UV} \leq z \leq z_{\rm IR}$ for smoothness and stability of the solution \footnote{A node in the solution $\bar X$ does not indicate an instability per se. However in this case it would happens in a region where the field $\bar X$ is small. Then the solution for the fluctuation of the field at zero mass will also have a node, which signals the presence of a mode with negative mass squared, i.e., an instability (see for example~\cite{Anguelova:2013tha})}.
Thus we may parametrize
\dis{
z_{\rm IR} \equiv z_{\rm UV} e^{(\pi - \beta)/\sqrt{\epsilon}} \label{mirb}
}
with $0< \alpha < \beta < \pi$, which reproduces the Miransky scaling if $\beta \ll 1$. 

The generation of the dynamical scale $m_{\rm dyn} \sim z_{\rm IR}^{-1}$ corresponds to stabilizing the radion in the 5D theory, for instance, by the Goldberger-Wise (GW) mechanism \cite{Goldberger:1999uk}. The GW mechanism can be implemented here by introducing boundary potentials for the bulk scalar field $X$.
The boundary potentials will also determine the parameters $\sigma$ and $\alpha$ as well as $z_{\rm IR}$ (thus, $\beta$). 
In this work we will not specify a radion stabilization mechanism. Instead we investigate the conditions for the parameters $\sigma$, $\alpha$ and $\beta$ needed to obtain a parametrically light dilaton and see its physical implications.

Depending on the values of $\alpha$ and $\beta$, we identify the following main scenarios:
\begin{itemize}
 \item[I] The standard scenario with Miransky scaling, where both $\alpha$ and $\beta$ are small: $\alpha \sim \beta \sim \sqrt{\epsilon}$. In this case, as we shall show below, we cannot find a parametrically light mode. 
 \item[II] A generic case where $\beta$ is not small, $\beta \sim 1$. In this case, interestingly, a light mode will be found assuming natural boundary conditions for the fluctuations. However, as seen from~\eqref{mirb}, this choice leads to a  non-standard scaling between the UV and IR parameters which deviates from the expected Miransky scaling.
 \item[II*] A subscenario where  $\beta \sim 1$ and also  $\alpha \sim 1$ but in such a way that the difference is small, $\beta-\alpha \ll 1$. For definiteness we shall assume the scaling $\beta -\alpha \sim \epsilon^{1/4}$, because it is singled out by the analysis of the condensate and the IR boundary condtions, as we will show below. 
\end{itemize}

Many earlier works analyze light scalars in nearly conformal holographic setups. These studies include examples in top-down constructions~\cite{Elander:2009pk,Elander:2010wd,Kutasov:2011fr,Anguelova:2012ka,Anguelova:2012mu,Elander:2017cle,Elander:2017hyr,Elander:2018gte} as well as  simple~\cite{Elander:2010wd,Haba:2010hu, Megias:2014iwa, Cox:2014zea} and somewhat more complex~\cite{Bellazzini:2013fga} bottom-up models.
In~\cite{Evans:2013vca,Alho:2013dka} the authors  find that a parametrically light state in a model of chiral symmetry breaking inspired by the D3-D7 gravity dual. 
In this model the gauge dynamics is put by hand by choosing the radial dependence of the quark mass which is interpreted as the renormalization group flow of the dimension of the quark bilinear.
There are also models which include 
 the flow from an asymptotically free theory to a (near) conformal IR or the consequent flow from the vicinity of the IR point to a QCD-like phase~\cite{Jarvinen:2011qe,Kutasov:2012uq,Goykhman:2012az,Pomarol:2019aae}.  We will discuss below how our findings compare to these earlier results.
 
\section{Parametrically light dilaton from the holographic model}

Let us consider 5D linear perturbation theory expanded around the AdS geometry (\ref{AdSm}) and bulk scalar background (\ref{X0}). 
For the perturbed metric, we may write in the Fefferman-Graham coordinate 
\dis{
ds^2 = \frac{L^2}{z^2}  \left[e^{2A(z)}(\eta_{\mu \nu} + h_{\mu \nu}) dx^\mu dx^\nu - (1+2 \phi) dz^2\right], \label{FG}
}
where $A(z)$, $h_{\mu \nu}(x, z)$, and $\phi(x, z)$ are small perturbations from the AdS$_5$ spacetime.
On the other hand, the scalar field perturbation will be denoted by $\chi(x,z)$, i.e.
\dis{
X(x, z) = \bar{X}(z) + \chi(x, z).
}

As the background is independent of 4D spacetime coordinates, it is enough to study the mass eigenmodes for the fluctuations, $(\square_4+m^2)f=0$, where $f$ represents any of the perturbations defined above.
Following the analysis of Ref.~\cite{Kiritsis:2006ua}, there are total 6 gauge-invariant degrees of freedom. 
For 4D massive modes\footnote{As shown in \cite{Kiritsis:2006ua}, there can be also 4D massless modes in the approximate $\textrm{AdS}_5$ geometry with an IR boundary with suitable boundary conditions. Those modes are projected out by applying boundary conditions similar to Eq. (\ref{suv}) and Eq. (\ref{sir}), i.e. $\Phi(x, z_{\rm UV})=0$ and $d\Phi(x, z)/dz |_{z= z_{\rm IR}}=0$ for a massless field $\Phi(x, z)$.}, five of them constitute a 4D transverse and traceless tensor  $h_{\mu \nu}^{TT}$, and the remaining one degree of freedom corresponds to a 4D scalar field $S$. 
The 4D scalar field is given by
\dis{
S = \psi - \frac{\chi}{Z}, 
}
where
\dis{
\psi \equiv \frac{1}{6} \left(h^\mu_\mu - \frac{\partial^\mu \partial^\nu h_{\mu \nu}}{m^2} \right), \label{psi}
}
and
\dis{
Z(z) \equiv \frac{\bar{X}’}{A’-1/z}.
} 
The field $S$ is a mixture of the radion field and $\chi$, which provides the dual description of the 4D dilaton 
arising from the gluon and quark condensations.

In linear perturbation theory, the equations of motion of $S$ is given by
 \dis{
 S'' + \left(3\frac{a’}{a} + 2 \frac{Z’}{Z} \right) S’ + m_S^2 S = 0\,, \label{seq}
 }
where the primes denote derivatives with respect to $z$.
 On the other hand, the graviton field $h_{\mu \nu}^{TT}$ obeys
  \dis{
 (h_{\mu \nu}^{TT})'' + 3\frac{a’}{a} (h_{\mu \nu}^{TT})’ + m_G^2 h_{\mu \nu}^{TT} = 0\,. \label{heq}
 }
For the approximate AdS background (\ref{warp}) and bulk scalar VEV (\ref{X0}),
 \bea 
\frac{a’}{a} &\simeq& -\frac{1}{z}\left[1 + {\cal O}\left(\bar{X}^2\right)\right], \label{a’a}\\
\frac{Z’}{Z}  &\simeq& \frac{2}{z}\Big[ 1 + {\cal O}\left(\bar{X}^2\right) +{\cal O} \left(\frac{\sqrt\epsilon}{\tan(\sqrt{\epsilon} \ln (z/z_{\rm UV})+ \alpha)}\right)\Big],  \nonumber \\
\label{Z’Z}
 \eea
 where the subleading corrections of order ${\cal O}(\bar{X}^2)$ are from the backreaction of the metric against a non-zero VEV of the $X$ field.
That correction depends on the value of $\sigma$. For the specific scenario II*, we will later show that $\sigma \sim z_{\rm IR}^{-2} \epsilon^{-1/4}$, so that the solution in Eq.~\eqref{X0} satisfies $\bar{X}(z) \ll 1$ except for a short region with $z \sim z_{\rm IR}$. We have studied numerically the effect of the ${\cal O}\left(\bar{X}^2\right)$ corrections in~\eqref{a’a} and~\eqref{Z’Z} for such a scenario. The effect turns out to be small and will not affect any of the discussion below. 
 In Eq. (\ref{Z’Z}) we have another correction which is proportional to $\sqrt{\epsilon}$. The correction can be sizable near $z\simeq z_{\rm UV}$ unless $\sqrt{\epsilon} \ll \alpha$ and also near $z \simeq z_{\rm IR}$ unless $\sqrt{\epsilon} \ll \beta-\alpha$. 
Thus let us assume 
 \dis{
 \sqrt{\epsilon} \ll \alpha \quad \textrm{and} \quad \sqrt{\epsilon} \ll \beta-\alpha \label{abcond}
 }
in order to ignore the correction terms. Later we will come back to discuss implications of this assumption for the existence of a light mode and the Miransky scaling. Notice that this assumption excludes the standard scenario I defined above.
  
Ignoring the subleading corrections in Eq. (\ref{a’a}) and Eq. (\ref{Z’Z}), solutions to the equations of motion Eq. (\ref{seq}) and Eq. (\ref{heq}) turn out to be
\bea
\!\! S(x, z)\!\! &\simeq&\!\! S_J (x) J_{0} (m_S z) + S_Y (x) Y_{0} (m_S z), \\
\!\!\! h_{\mu \nu}^{TT}(x, z) \!\! & \simeq & \!\! h_J (x) z^2 J_2(m_G z) + h_Y (x) z^2 Y_2 (m_G z),  \label{h5D}
\eea
where $J_n (z)$ and $Y_n(z)$ are the Bessel functions of the first kind and the second kind, respectively. The functions $S_{J/Y}(x)$ and $h_{J/Y}(x)$ are plane waves with the value of the mass given by $m_S$ and $m_G$, respectively.

Let us first analyze the KK modes of the $S$ field. 
For the UV normalizability and the IR regularity of physical modes, we require boundary conditions
\bea
 S(x, z_{\rm UV}) &=& 0, \label{suv}\\
 S’(x, z_{\rm IR}) &=& 0, \label{sir}
\eea
 which lead to
 \dis{
- \frac{S_Y(x)}{S_J (x)} = \frac{J_0 (m_S z_{\rm UV}) }{Y_0 ( m_S z_{\rm UV}) }=\frac{J_1 (m_S z_{\rm IR}) }{Y_1 ( m_S z_{\rm IR})}. \label{sbc}
 }
Thus $S(x, z)$ can be expanded in terms of normalizable modes as follows.
\dis{
 S(x, z) = \sum_{n=0}^{\infty} S^{(n)}(x) f_S^{(n)}(z)
}
with
\dis{
f_S^{(n)}(z) = \frac{1}{N_S^{(n)}}\left( J_0(m_S^{(n)} z) - \frac{J_0(m_S^{(n)} z_{\rm UV})}{Y_0(m_S^{(n)} z_{\rm UV})} Y_0 (m_S^{(n)} z)\right)
} 
where $m_S^{(n)}$ denotes a solution of $m_S$ which satisfies Eq. (\ref{sbc}), corresponding to the mass of the KK mode $S^{(n)}(x)$, and $N_S^{(n)}$ is a certain normalization factor responsible
for the canonical normalization of $S^{(n)}(x)$. 

From Eq. (\ref{sbc}), one can derive an approximate analytic expression for the KK mass $m_S^{(n)}$.
For the lightest mode $S^{(0)}(x)$, we find 
 \dis{
 m_S^{(0)} &\simeq \frac{\sqrt{2}}{z_{\rm IR}} \frac{1}{\sqrt{\ln (z_{\rm IR}/z_{\rm UV})}} = \frac{\epsilon^{1/4}}{z_{\rm IR}} \sqrt{\frac{2}{\pi -\beta}} \ll m_{\rm dyn}\,. \label{mD}
 }
 On the other hand, the heavier modes $S^{(n)}(x)$, with $n= 1, 2, \cdots$, are given by the zeroes of the Bessel $J_1$ function and satisfy approximately
 \dis{ \label{SKKmodes}
 m_S^{(n)} \approx \frac{\pi}{z_{\rm IR}} \left(n + \frac{1}{4} \right) \gtrsim m_{\rm dyn}\,.
 }
 This approximate expression is more accurate for large $n \gg 1$.
Therefore, one can see that the lightest mode $S^{(0)}(x)$ is parametrically lighter than the dynamical scale $m_{\rm dyn} \sim z_{\rm IR}^{-1}$, while the other modes have masses around or above the dynamical scale.
The lightest mode would be identified as the holographic description of the dilaton state of the boundary 4D theory. The parametrically light mass of the dilaton indicates that the dilaton decay constant is parametrically larger than the dynamical scale. This is expected for the gauge theories at the conformal edge, because the dilaton decay constant is the vacuum expectation value (vev) of the dilaton field where the scale symmetry is spontaneously broken, while the dynamical mass characterizes the size of explicit breaking of scale symmetry that can be made arbitrarily smaller than the dilaton decay constant~\footnote{The dilaton, a state created by the dilation current out of the vacuum, remains massless in the limit the dynamical mass or the explicit breaking of the scale symmetry vanishes. The well-known example is the scale-symmetry breaking by the Coleman-Weinberg mechanism where the scalar mass $m\sim\lambda v\to0$ for $\lambda\to0$, while the vev $v$ remains finite .}. 

One may wonder whether such light mode does exist only for the scalar field as expected from the boundary 4D theory. 
To check this, we also examine the KK mass spectrum of the graviton field $h_{\mu \nu}^{TT}(x, z)$. 
Imposing the boundary conditions for the UV normalizability and the IR regularity of physical modes, that is
 \bea
 h_{\mu \nu}^{TT}(z_{\rm UV}) &=& 0,\\
 h_{\mu \nu}^{TT’}(z_{\rm IR}) &=& 0,
 \eea
 it comes out that
 \dis{ \label{Gcond}
\frac{J_2(m_G z_{\rm UV})}{Y_2(m_G z_{\rm UV})}=\frac{J_1(m_G z_{\rm IR})}{Y_1(m_G z_{\rm IR})}.
 }
From this condition, we obtain the KK mass spectrum for the graviton field as follows: 
\dis{
m_G^{(n)} \approx \frac{\pi}{z_{\rm IR}} \left( n+ \frac{1}{4}\right) \gtrsim m_{\rm dyn}\,, \label{grav}
}
where $n= 1, 2, \cdots$, which is more accurate for large $n$.  
Even though the condition for the heavy modes is the same as for the scalar states in~\eqref{SKKmodes}, by studying the condition~\eqref{Gcond} at small mass we find that
there is no non-scalar state parametrically lighter than the dynamical scale $m_{\rm dyn} \sim z_{\rm IR}^{-1}$.

Let us finally discuss the relation between the existence of a light scalar mode and our assumption on the parameters $\alpha$ and $\beta$, Eq. (\ref{abcond}) and $\bar{X}(z) \ll 1$ to ignore the subleading corrections in Eq. (\ref{a’a}) and Eq. (\ref{Z’Z}). As we have shown above, the graviton field does not have a light mode, while the scalar field does. Their approximated field equations of motion are from Eq. (\ref{seq}) and Eq. (\ref{heq})
\bea
S'' + \frac{1}{z} S' + m_S^2 S &=& 0, \label{seq2} \\
(h_{\mu \nu}^{TT})'' - \frac{3}{z} (h_{\mu \nu}^{TT})' + m_G^2 h_{\mu \nu}^{TT} &=& 0, \label{heq2}
\eea
by Eq. (\ref{a’a}) and Eq. (\ref{Z’Z}) ignoring the subleading corrections.
Thus one can observe that the difference between them is originated from the coefficients of the first derivative terms in the field equations. It means that 
the coefficient of the first derivative term is important for the existence of a light mode. If we allow a significant correction from the subleading terms in Eq. (\ref{a’a}) and Eq. (\ref{Z’Z}), it will change the coefficient of the $S'/z$ term in Eq. (\ref{seq2}), and it turns out to spoil a parametrically light mass of the lightest scalar mode.
Although we have not rigorously proven that Eq. (\ref{abcond}) and $\bar{X} \ll 1$ is a necessary condition to get a light mode, we find that it is a sufficient condition and implies a non-standard scaling between the UV and IR parameters according to Eq. (\ref{mirb}) deviating from the Miransky scaling.


\section{Symmetric Relations}

In this section, we will verify that conformal symmetric relations valid in the 4D gauge theory are established in our holographic model. 
By the dilatation symmetry, it can be shown that there is an associated Ward-Takahashi identity in low momentum limit $q_\mu \rightarrow 0$ but $q^2 \gg m_D^2 \approx 0$ (with $m_D$ dilaton mass) \cite{Choi:2012kx}: 
\dis{
\lim_{q_\mu \rightarrow 0} \int d^4 x \,e^{iq\cdot x} \langle 0 | T \theta^\mu_\mu(x) \theta^\nu_\nu(0) | 0 \rangle = 4 i  \langle 0 | \theta^\mu_\mu | 0 \rangle. \label{wt}
}
Here $\theta^\mu_\nu$ is the energy-momentum tensor of the 4D gauge theory. 
On the other hand, the Partially Conserved Dilatation Current (PCDC)  relation is often stated as \cite{Choi:2012kx, Kawaguchi:2020kce}
\dis{
f_D^2 m_D^2 = -4 \langle 0 | \theta^\mu_\mu | 0 \rangle, \label{pcdc}
}
where $f_D$ is the dilaton decay constant.

Let us first examine the Ward-Takahashi identity Eq. (\ref{wt}) by holographic computation. 
The correlation functions can be calculated in the 5D dual theory following the prescription from the AdS-CFT correspondence:
\bea
\langle {\theta^{\mu}_\mu}(x) \rangle &=& -\left.\frac{\delta W}{\delta \psi(x)} \right|_{\psi=0}\,, \label{1pt} \\
\langle {\theta^{\mu}_\mu}(x) {\theta^{\nu}_\nu}(0) \rangle &=& -i \left.\frac{\delta^2 W}{\delta \psi(x) \delta \psi(0)}  \right|_{\psi=0}\,, \label{2pt}
\eea
where $W$ is the generating functional of the connected correlators of 4D boundary theory, obtained from the 5D holographic action on shell, which we will give explicitly below, and $\psi$ is defined in Eq. (\ref{psi}).

The relevant part for the two-point correlation function can be obtained from the second derivative of $W$ with respect to the $S$ field using $\delta S(x_1) / \delta \psi (x_2) = \delta^{4}(x_1-x_2)$.
 The 5D holographic action for the $S$ field is given by
\dis{
{\cal S} = \frac{1}{\kappa_5^2}\int d^4 x dz \sqrt{g} \frac{1}{2} Z^2 g^{MN} \partial_M S \partial_N S\,.
}
By integrating by parts and applying the equation of motion (\ref{seq}), this action gives rise to the following 4D generating functional,
\dis{
W[S] =\frac{1}{\kappa_5^2}\int d^4 x \left.\left(-\frac{1}{2}a^3 Z^2 S \partial_z S\right) \right|_{z=z_{\rm UV}}.
}

For the source field $S(x,z)$, one can write it as a linear combination of a non-normalizable mode and a normalizable mode as following in the 4D Fourier space:
\dis{
\tilde{S}(q^\mu,z) &=\Big[  \frac{J_0(q z)}{J_0(q z_{\rm UV})} \\
&+ c_2(q) \left(Y_0(q z) -\frac{Y_0(q z_{\rm UV})}{J_0(q z_{\rm UV})} J_0 (q z) \right) \Big] \tilde{S} (q^\mu)\,,
}
where $q\equiv\sqrt{q^2}$. The coefficient $c_2(q)$ is  
to be determined by a boundary condition. 
Note that the source $\tilde{S}(q^\mu, z)$ is normalized to be $\tilde{S}(q^\mu, z_{\rm UV})  = \tilde{S}(q^\mu)$.
Imposing the IR regularity condition for $\tilde{S}(q^\mu, z)$, i.e.
\dis{
\tilde{S}’(q^\mu, z_{\rm IR}) =0,
}
we get
\bea
c_2(q) &=& \left(Y_0(q z_{\rm UV}) -\frac{Y_1(q z_{\rm IR})}{J_1(q z_{\rm IR})} J_0 (q z_{\rm UV})\right)^{-1} \\
&\simeq& -\frac{\pi}{4}  \frac{q^2 m_D^2 }{q^2 - m_D^2}z_{\rm IR}^2  \quad\textrm{for}~ q^2 \approx m_D^2\,,
\eea
where we identify the dilaton mass $m_D \equiv m_S^{(0)}$, and the last line is obtained by expanding $c_2(q)$ around the pole at $q^2 = (m_S^{(0)})^2$.
The two point correlation function then comes out as
\dis{
\int d^4 x \,&e^{iq\cdot x} \langle 0 | T \theta^\mu_\mu(x) \theta^\nu_\nu(0) | 0 \rangle = i \frac{8}{\pi} M_5^3 L^3 c_2(q) \sigma^2 \sin^2 \alpha  
\\
&\simeq -2i M_5^3 L^3 \frac{q^2   m_D^2 }{q^2 - m_D^2} \sigma^2 z_{\rm IR}^2 \sin^2 \alpha \quad\textrm{for}~ q^2 \approx m_D^2\,.
}
Taking the limit $q^\mu \rightarrow 0$ but $q^2 \gg m_D^2 \approx 0$, we find
\dis{
\lim_{q_\mu\rightarrow 0} \int d^4 x \,&e^{iq\cdot x} \langle 0 | T \theta^\mu_\mu(x) \theta^\nu_\nu(0) | 0 \rangle \\
&\simeq -2i M_5^3 L^3 m_D^2 \sigma^2 z_{\rm IR}^2 \sin^2 \alpha\,. \label{2ptr}
}

Next let us compute the one-point function from the holography. The first order perturbation around the background solution is always a boundary term. 
For the gauge choice $\phi = 0$ in Eq. (\ref{FG}), this first order action simplifies to
\dis{
W^{(1)} = \frac{L^3}{2\kappa_5^2} \frac{e^{4A(z)}}{z^4} \int d^4x \Big[&24 (1-zA’(z)) \psi(x, z)  \\
&+ 2z \bar{X}’(z) \chi(x, z) \Big]_{z=z_{\rm UV}}.
}
The action $W^{(1)}$ is UV divergent, and holographic renormalization is required. It is well known how generic Einstein-scalar actions are renormalized (see~\cite{Papadimitriou:2011qb} for a thorough discussion). In our case, however, the violation of the BF bound leads to a minor ambiguity in the finite counter terms. We write the counter terms as
\dis{
W_{\rm ct} = -\frac{1}{2\kappa_5^2} \int d^4x \sqrt{-\gamma}\Big[ \frac{6}{L} + \frac{2 + c\sqrt{\epsilon}}{L} X^2 \Big]_{z = z_{\rm UV}} \label{counter}
} 
where $\gamma$ is the determinant of the induced metric $\gamma_{\mu \nu}(x)$ at the UV boundary, 
and $c$ is a coefficient that represents the ambiguity of the holographic renormalization in the case $0< \epsilon \ll 1$. For the QCD-like case $-1<\epsilon <0$, $c=i$. 
Including the counter terms, the renormalized first order action is
\dis{
W^{(1)}_{\rm ren} &= \frac{L^3}{2\kappa_5^2} \frac{e^{4A(z)}}{z^4} \int d^4x \\
&\Big[-4 \big((2+c\sqrt{\epsilon})\bar{X}^2(z) 
+ 6zA’(z)\big) \psi(x, z)  \\
&+ 2\left(z \bar{X}’(z) - (2+c\sqrt{\epsilon}) \bar{X}(z)\right) \chi(x, z) \Big]_{z=z_{\rm UV}}. \label{W1ren}
}
The perturbed metric $A(z)$ can be shown to obey the following equation of motion,
\dis{
3 z^2 A'' + 3 z A’ + (z\bar{X}’)^2 =0.
}
For the background solution (\ref{X0}) for $X$, $A(z)$ is thus given by
\dis{
A(z) = -\frac{1}{24} \sigma^2 z^4 \left[1-\cos 2 \left(\sqrt{\epsilon} \ln \left(\frac{z}{z_{\rm UV}} \right) + \alpha \right) + \frac{\epsilon}{4}\right]. \label{Az}
}
Inserting the solutions Eq. (\ref{Az}) and Eq. (\ref{X0}) into the action (\ref{W1ren}), one can get
the one-point function by Eq. (\ref{1pt})
\bea \label{1ptr1}
\langle {\theta^{\mu}_\mu}(x) \rangle &\simeq& -2 M_5^3 L^3  \sigma^2 \sqrt{\epsilon}  (\sin 2 \alpha-2c\sin^2 \alpha) \\
&\simeq& - M_5^3 L^3 \sigma^2 m_D^2 z_{\rm IR}^2 (\pi-\beta)(\sin 2 \alpha-2c\sin^2 \alpha). \nonumber \\ \label{1ptr}
\eea
For the last line, we have used Eq. (\ref{mD}).

From Eq. (\ref{2ptr}) and Eq. (\ref{1ptr}), we find that the Ward identity (\ref{wt}) is satisfied 
if the IR parameter $\beta$ is related to the counterterm $c$ and the UV parameter $\alpha$ as
\dis{
\beta=\pi+\frac{1}{4(c-\cot \alpha)}.
}
Because of the cotangent term, we would have that $|c|\gg 1$ in scenario I where $\alpha \sim \beta \sim \sqrt{\epsilon}$. This is undesired as the expected ambiguity in the counterterm~\eqref{counter} is $c \sim 1$. Recall however that scenario I was already excluded because it does not produce a parametrically light mode. In the scenarios II and II*, where we can have $\alpha \sim 1$, this issue does not appear. 

On the other hand, from Eq. (\ref{pcdc}), the dilaton decay constant is identified as 
\dis{
f_D^2 \simeq 2M_5^3 L^3 \sigma^2 z_{\rm IR}^2 \sin^2 \alpha.
}
According to the Miransky and Gusynin \cite{Miransky:1989qc},
\dis{
 \langle \theta^\mu_\mu (x) \rangle \sim m_{\rm dyn}^4 \sim z_{\rm IR}^{-4}.
 }
From Eq.  (\ref{1ptr1}) we see that (assuming $\alpha \sim 1$ so that there is no issue with the Ward identity), this scaling is obtained if
 \dis{ \label{sigmasc}
\sigma \sim z_{\rm IR}^{-2} \epsilon^{-1/4}.
}
Interestingly, this is consistent with a natural\footnote{This condition has been seen to arise effectively from a complex IR dynamics, for example, in~\cite{Jarvinen:2011qe,Jarvinen:2015ofa}.} IR boundary condition $\bar X(z_{\rm IR}) \sim 1$ for scenario II*: by using~\eqref{sigmasc} and~\eqref{X0}, we find
\dis{
 X(z_{\rm IR}) \sim \sigma z_{\rm IR}^2 \sin\left(\pi +\alpha - \beta \right) \sim 1 .
}
Notice that this requires the specific scaling, $\beta-\alpha \sim \epsilon^{1/4}$, assumed for scenario II*.

\section{Conclusions} \label{sec:conc}
We study a bottom-up holographic model  for gauge theories near the conformal edge, constructed in a slice of the 5D anti-de Sitter space. The theory breaks scale symmetry spontaneously when the quark bilinear develops a condensate, generating the infrared scale, $m_{\rm dyn}$, which is represented in the holographic model by the infrared cutoff of the AdS slice, $z_{\rm IR}$.
By explicitly analyzing the spectra of the composite states, choosing the parameters of the model such that the back reaction to the bulk geometry is negligible, we find there is a unique scalar state whose mass is parametrically lighter than all other states, whose masses are around $m_{\rm dyn}$ or higher. The lightest scalar is identified as the dilaton, the Nambu-Goldstone boson associated with the scale symmetry, by showing that it saturates at low energies the anomalous Ward identity of dilatation currents.  The dilaton mass vanishes as the dimension of the quark bilinear approaches the Breitenlohner-Freedman bound, which shows that the dilaton decay constant is parametrically larger than the infrared scale, having the walking behavior of the theory between two scales, widely separated by the Miransky scaling. 

Our model adds to the wide literature where parametrically light modes have been analyzed by using holography. The results from these studies vary: Several specific models do show signs of light scalars, which may even be parametrically light (see, e.g.~\cite{Haba:2010hu,Evans:2013vca,Alho:2013dka,Elander:2017cle,Elander:2017hyr}). 
Models with more complex structure often have scalar which are relatively light compared to the other mesons, but not parametrically light~\cite{Kutasov:2011fr,Kutasov:2012uq,Goykhman:2012az,Arean:2012mq,Arean:2013tja,Elander:2018gte,Pomarol:2019aae}. In this article we started by analyzing a class of simple models, with UV and IR boundary conditions parametrized, among other things, by the real parameters $\alpha$ and $\beta$. We identified a subclass of models (scenario II) which shows a parametrically light states whose mass could be found analytically in~\eqref{mD}. Therefore, while our work is not generic enough to capture the precise dynamics of the various examples considered in the literature, it should anyhow be seen as a step towards the classification of light scalars in nearly conformal models. 

A possible improvement of our study would be to include an explicit realization of the UV and IR stabilization mechanism giving rise to the parameters $\alpha$ and $\beta$ or an explanation for the origin of the IR boundary conditions of the fluctuations. In future, we are planning to develop the model in this direction by including similar, and equally general, mechanisms directly in the holographic setup.


\acknowledgments

This work was supported by the National Research Foundation of Korea (NRF) grants funded by the Korean government (MSIT) (grant numbers 2021R1A4A5031460 (DKH) and 2021R1A2C1010834 (MJ)), Basic Science Research Program through the National Research Foundation of Korea (NRF) funded by the Ministry of Education (NRF- 2017R1D1A1B06033701) (DKH),
and was also supported by IBS under the project code, IBS-R018-D1 (SHI). JCR and MJ have been
supported by an appointment to the JRG Program at the APCTP through the
Science and Technology Promotion Fund and Lottery Fund of the Korean
Government. JCR and MJ have also been supported by the Korean Local
Governments -- Gyeong\-sang\-buk-do Province and Pohang City. 
This work benefited from discussions during the APCTP focus program ``QCD and gauge/gravity duality''.

%



\begin{thebibliography}{99}
\bibitem{Hong:2017smd}
D.~K.~Hong,
JHEP \textbf{02}, 102 (2018)
doi:10.1007/JHEP02(2018)102
[arXiv:1703.05081 [hep-ph]].

\bibitem{Eroncel:2018dkg}
C.~Er\"oncel, J.~Hubisz and G.~Rigo,
JHEP \textbf{03}, 046 (2019)
doi:10.1007/JHEP03(2019)046
[arXiv:1804.00004 [hep-ph]].

\bibitem{Choi:2011fy}
K.~Y.~Choi, D.~K.~Hong and S.~Matsuzaki,
Phys. Lett. B \textbf{706}, 183-187 (2011)
doi:10.1016/j.physletb.2011.11.013
[arXiv:1101.5326 [hep-ph]].
 
\bibitem{Choi:2012kx}
K.~Y.~Choi, D.~K.~Hong and S.~Matsuzaki,
JHEP \textbf{12}, 059 (2012)
doi:10.1007/JHEP12(2012)059
[arXiv:1201.4988 [hep-ph]].
 

\bibitem{Gorbenko:2018ncu}
V.~Gorbenko, S.~Rychkov and B.~Zan,
JHEP \textbf{10}, 108 (2018)
doi:10.1007/JHEP10(2018)108
[arXiv:1807.11512 [hep-th]].
 

\bibitem{Kaplan:2009kr}
D.~B.~Kaplan, J.~W.~Lee, D.~T.~Son and M.~A.~Stephanov,
Phys. Rev. D \textbf{80}, 125005 (2009)
[arXiv:0905.4752 [hep-th]].

\bibitem{Faedo:2019nxw}
A.~F.~Faedo, C.~Hoyos, D.~Mateos and J.~G.~Subils,
Phys. Rev. Lett. \textbf{124}, no.16, 161601 (2020)
doi:10.1103/PhysRevLett.124.161601
[arXiv:1909.04008 [hep-th]].

\bibitem{Faedo:2021ksi}
A.~F.~Faedo, C.~Hoyos, D.~Mateos and J.~G.~Subils,
JHEP \textbf{10}, 246 (2021)
doi:10.1007/JHEP10(2021)246
[arXiv:2106.01802 [hep-th]].

\bibitem{Alanen:2010tg}
J.~Alanen, K.~Kajantie and K.~Tuominen,
Phys. Rev. D \textbf{82}, 055024 (2010)
doi:10.1103/PhysRevD.82.055024
[arXiv:1003.5499 [hep-ph]].
 
\bibitem{Maldacena:1997re}
J.~M.~Maldacena,
Adv. Theor. Math. Phys. \textbf{2}, 231-252 (1998)
[arXiv:hep-th/9711200 [hep-th]].


\bibitem{Appelquist:1996dq}
T.~Appelquist, J.~Terning and L.~C.~R.~Wijewardhana,
Phys. Rev. Lett. \textbf{77}, 1214-1217 (1996)
doi:10.1103/PhysRevLett.77.1214
[arXiv:hep-ph/9602385 [hep-ph]].




\bibitem{Hong:2006si}
D.~K.~Hong and H.~U.~Yee,
Phys. Rev. D \textbf{74}, 015011 (2006)
doi:10.1103/PhysRevD.74.015011
[arXiv:hep-ph/0602177 [hep-ph]].

\bibitem{Anguelova:2013tha}
L.~Anguelova, P.~Suranyi and L.~C.~R.~Wijewardhana,
Nucl. Phys. B \textbf{881}, 309-326 (2014)
doi:10.1016/j.nuclphysb.2014.02.010
[arXiv:1306.1981 [hep-th]].


\bibitem{Goldberger:1999uk}
W.~D.~Goldberger and M.~B.~Wise,
Phys. Rev. Lett. \textbf{83}, 4922-4925 (1999) 
doi:10.1103/PhysRevLett.83.4922
[arXiv:hep-ph/9907447 [hep-ph]].


\bibitem{Elander:2009pk}
D.~Elander, C.~Nunez and M.~Piai,
Phys. Lett. B \textbf{686}, 64-67 (2010)
doi:10.1016/j.physletb.2010.02.023
[arXiv:0908.2808 [hep-th]].

\bibitem{Elander:2010wd}
D.~Elander and M.~Piai,
JHEP \textbf{01}, 026 (2011)
doi:10.1007/JHEP01(2011)026
[arXiv:1010.1964 [hep-th]].


\bibitem{Kutasov:2011fr}
D.~Kutasov, J.~Lin and A.~Parnachev,
Nucl. Phys. B \textbf{858}, 155-195 (2012)
doi:10.1016/j.nuclphysb.2012.01.004
[arXiv:1107.2324 [hep-th]].

 
\bibitem{Anguelova:2012ka}
L.~Anguelova, P.~Suranyi and L.~C.~R.~Wijewardhana,
Nucl. Phys. B \textbf{862}, 671-690 (2012)
doi:10.1016/j.nuclphysb.2012.05.005
[arXiv:1203.1968 [hep-th]].

\bibitem{Anguelova:2012mu}
L.~Anguelova, P.~Suranyi and L.~C.~R.~Wijewardhana,
JHEP \textbf{05}, 003 (2013)
doi:10.1007/JHEP05(2013)003
[arXiv:1212.1176 [hep-th]].

\bibitem{Elander:2017cle}
D.~Elander and M.~Piai,
Phys. Lett. B \textbf{772}, 110-114 (2017)
doi:10.1016/j.physletb.2017.06.035
[arXiv:1703.09205 [hep-th]].

\bibitem{Elander:2017hyr}
D.~Elander and M.~Piai,
JHEP \textbf{06}, 003 (2017)
doi:10.1007/JHEP06(2017)003
[arXiv:1703.10158 [hep-th]].

\bibitem{Elander:2018gte}
D.~Elander, A.~F.~Faedo, D.~Mateos, D.~Pravos and J.~G.~Subils,
JHEP \textbf{05}, 175 (2019)
doi:10.1007/JHEP05(2019)175
[arXiv:1810.04656 [hep-th]].

\bibitem{Haba:2010hu}
K.~Haba, S.~Matsuzaki and K.~Yamawaki,
Phys. Rev. D \textbf{82}, 055007 (2010)
doi:10.1103/PhysRevD.82.055007
[arXiv:1006.2526 [hep-ph]].

\bibitem{Megias:2014iwa}
E.~Megias and O.~Pujolas,
JHEP \textbf{08}, 081 (2014)
doi:10.1007/JHEP08(2014)081
[arXiv:1401.4998 [hep-th]].

\bibitem{Cox:2014zea}
P.~Cox and T.~Gherghetta,
JHEP \textbf{02}, 006 (2015)
doi:10.1007/JHEP02(2015)006
[arXiv:1411.1732 [hep-th]].

\bibitem{Bellazzini:2013fga}
B.~Bellazzini, C.~Csaki, J.~Hubisz, J.~Serra and J.~Terning,
Eur. Phys. J. C \textbf{74}, 2790 (2014)
doi:10.1140/epjc/s10052-014-2790-x
[arXiv:1305.3919 [hep-th]].

\bibitem{Evans:2013vca}
N.~Evans and K.~Tuominen,
Phys. Rev. D \textbf{87}, no.8, 086003 (2013)
doi:10.1103/PhysRevD.87.086003
[arXiv:1302.4553 [hep-ph]].

\bibitem{Alho:2013dka}
T.~Alho, N.~Evans and K.~Tuominen,
Phys. Rev. D \textbf{88}, 105016 (2013)
doi:10.1103/PhysRevD.88.105016
[arXiv:1307.4896 [hep-ph]].


\bibitem{Jarvinen:2011qe}
M.~Jarvinen and E.~Kiritsis,
JHEP \textbf{03}, 002 (2012)
doi:10.1007/JHEP03(2012)002
[arXiv:1112.1261 [hep-ph]].

\bibitem{Kutasov:2012uq}
D.~Kutasov, J.~Lin and A.~Parnachev,
Nucl. Phys. B \textbf{863}, 361-397 (2012)
doi:10.1016/j.nuclphysb.2012.05.025
[arXiv:1201.4123 [hep-th]].

\bibitem{Goykhman:2012az}
M.~Goykhman and A.~Parnachev,
Phys. Rev. D \textbf{87}, no.2, 026007 (2013)
doi:10.1103/PhysRevD.87.026007
[arXiv:1211.0482 [hep-th]].

\bibitem{Pomarol:2019aae}
A.~Pomarol, O.~Pujolas and L.~Salas,
JHEP \textbf{10}, 202 (2019)
doi:10.1007/JHEP10(2019)202
[arXiv:1905.02653 [hep-th]].


\bibitem{Kiritsis:2006ua}
E.~Kiritsis and F.~Nitti,
Nucl. Phys. B \textbf{772}, 67-102 (2007)
doi:10.1016/j.nuclphysb.2007.02.024
[arXiv:hep-th/0611344 [hep-th]].



\bibitem{Kawaguchi:2020kce}
M.~Kawaguchi, S.~Matsuzaki and X.~G.~Huang,
JHEP \textbf{10}, 017 (2020)
doi:10.1007/JHEP10(2020)017
[arXiv:2007.00915 [hep-ph]].

\bibitem{Papadimitriou:2011qb}
I.~Papadimitriou,
JHEP \textbf{08}, 119 (2011)
doi:10.1007/JHEP08(2011)119
[arXiv:1106.4826 [hep-th]].

\bibitem{Miransky:1989qc}
V.~A.~Miransky and V.~P.~Gusynin,
Prog. Theor. Phys. \textbf{81}, 426-450 (1989)
doi:10.1143/PTP.81.426

\bibitem{Jarvinen:2015ofa}
M.~Jarvinen,
JHEP \textbf{07}, 033 (2015)
doi:10.1007/JHEP07(2015)033
[arXiv:1501.07272 [hep-ph]].

\bibitem{Arean:2012mq}
D.~Arean, I.~Iatrakis, M.~J\"arvinen and E.~Kiritsis,
Phys. Lett. B \textbf{720}, 219-223 (2013)
doi:10.1016/j.physletb.2013.01.070
[arXiv:1211.6125 [hep-ph]].

\bibitem{Arean:2013tja}
D.~Are\'an, I.~Iatrakis, M.~J\"arvinen and E.~Kiritsis,
JHEP \textbf{11}, 068 (2013)
doi:10.1007/JHEP11(2013)068
[arXiv:1309.2286 [hep-ph]].

 
 \end{thebibliography}
 \end{document}